\documentclass[aps,pra,10pt,superscriptaddress,nofootinbib,longbibliography,twocolumn]{revtex4-1}

\usepackage[utf8]{inputenc}
\usepackage{amsfonts}
\usepackage{amssymb}
\usepackage{pifont}
\usepackage{physics}
\usepackage{graphicx}
\usepackage{overpic}
\usepackage{color}
\usepackage[dvipsnames]{xcolor}
\usepackage[hidelinks]{hyperref}
\usepackage{times}
\usepackage{comment}

\usepackage[T1]{fontenc} 
\usepackage{booktabs}
\usepackage{float}

\usepackage{soul}

\usepackage{orcidlink}

\newcommand{\MJ}{\mathrm{MJ}}
\newcommand{\KJ}{\mathrm{KJ}}
\newcommand{\kW}{\mathrm{kW}}
\newcommand{\kWh}{\mathrm{kWh}}
\newcommand{\Wh}{\mathrm{Wh}}
\newcommand{\kWs}{\mathrm{kWs}}
\newcommand{\MFLOPSW}{\mathrm{M \; FLOPS} / \mathrm{W}}
\newcommand{\GFLOPSW}{\mathrm{G \; FLOPS} / \mathrm{W}}
\newcommand{\FLOP}{\mathrm{FLOP}}

\newcommand{\mpsoracle}{mps-oracle}

\newcommand{\efep}{EFEP}

\newcommand{\egate}{E_{\rm \scriptscriptstyle Gate}}
\newcommand{\erydalgo}{E_{\rm \scriptscriptstyle Algo}^{\rm \scriptscriptstyle Rydberg}}
\newcommand{\estartup}{E_{\rm \scriptscriptstyle Startup}}
\newcommand{\eload}{E_{\rm \scriptscriptstyle Load}}
\newcommand{\emeas}{E_{\rm \scriptscriptstyle Measure}}
\newcommand{\ecompile}{E_{\rm \scriptscriptstyle Compile}}

\newcommand{\eaqtalgo}{E_{\rm \scriptscriptstyle Algo}^{\rm \scriptscriptstyle AQT}}

\newcommand{\mqft}{m_{\rm \scriptscriptstyle QFT}}
\newcommand{\mrnd}{m_{\rm \scriptscriptstyle R}}

\newcommand{\orciddaniel}{\orcidlink{0000-0001-7658-3546}}
\newcommand{\orcidsimone}{\orcidlink{0000-0002-8882-2169}}

\newcommand{\uulm}{Institute for Complex Quantum Systems, 
Ulm University, Albert-Einstein-Allee 11, 89069 Ulm, Germany}
\newcommand{\paddip}{Dipartimento di Fisica e Astronomia "G. Galilei" \& Padua Quantum Technologies Research Center,
 Universit{\`a} degli Studi di Padova, Italy I-35131, Padova, Italy}
\newcommand{\padinfn}{INFN, Sezione di Padova, via Marzolo 8, I-35131,
  Padova, Italy}

\begin{document}

\title{Is quantum computing green? An estimate for an energy-efficiency
  quantum advantage}

\author{Daniel Jaschke\orciddaniel}
\affiliation{\uulm}
\affiliation{\paddip}
\affiliation{\padinfn}

\author{Simone Montangero\orcidsimone}
\affiliation{\uulm}
\affiliation{\paddip}
\affiliation{\padinfn}

\date{\today}


\begin{abstract}

  The quantum advantage threshold determines when a quantum processing unit (QPU) is more efficient with respect to classical computing hardware in terms of algorithmic complexity.
  The ``green" quantum advantage threshold -- based on a comparison of  energetic efficiency between the two -- is going to play a fundamental role in the comparison between quantum and classical hardware. Indeed, its characterization  would enable better decisions on energy-saving strategies, e.g. for distributing the workload in hybrid quantum-classical algorithms. Here, we show that the green quantum advantage threshold crucially depends on (i)~the quality of the experimental quantum gates and (ii)~the  entanglement generated in the QPU. Indeed, for NISQ hardware and algorithms requiring a moderate amount of entanglement, a classical tensor network emulation can be more energy-efficient at equal final state fidelity than quantum computation. We compute the green quantum advantage threshold for a few paradigmatic examples in terms of algorithms and hardware platforms, and identify algorithms with a power-law decay of singular values of bipartitions -- with power-law exponent $\alpha \lesssim 1$ -- as the green quantum advantage threshold in the near future.

\end{abstract}

\maketitle


The exascale computing era is making the energy consumption of high-performance computing (HPC) a crucial aspect of modern facilities: energy-saving optimizations in 
the computational cluster workflow play a fundamental role~\cite{Mammela2012,Lim2015,Agosta2022}.
The rise of graphics processing units (GPU) for HPC applications demonstrated
the benefit of specialized hardware accelerating certain tasks, e.g.,
for machine learning applications~\cite{Kurth2018,Laanait2019}. The integration of quantum computers in HPC environments can potentially provide another accelerating hardware serving as quantum processing units (QPU)~\cite{Suchara2018,Villalonga2020,Johansson2021,Auffeves2021}. Quantum processing units might be beneficial either for speeding up the overall computation - exploiting the quantum advantage - or drastically reducing the energy consumption of a task, i.e., displaying a green quantum advantage.
Both advantages influence the decisions of how to split the workload in a hybrid
quantum-classical algorithm or infrastructure and can occur at the same time.
The estimation of resource constraints for quantum computing from compilation to running the hardware has already attracted interest, even beyond the noisy intermediate-scale quantum (NISQ) regime for fault-tolerant applications~\cite{Preskill2018,FellousAsiani2021,Paler2022}.

The quantum advantage is expected to induce also a green quantum advantage because
the beneficial scaling in time of a Bounded-error Quantum Polynomial time (BPQ) problem on a quantum machine versus an NP
problem on a classical computer will dominate with large enough problem sizes~\cite{Bernstein1993}. But there
remain two questions to be determined: do algorithms exist that have no quantum advantage
but nonetheless show a green quantum advantage and to which extent the beneficial scaling of BPQ
versus NP is observable with current system sizes. After recent breakthroughs in the QPU
development which shifted the border for a quantum advantage in favor of QPUs~\cite{Arute2019,Zhong2020},
classical algorithms and tensor network simulations have followed up on the results
and improved the classical simulation problem with novel
approaches~\cite{Liu2021,Bulmer2022}. The benefit of tensor network methods is the ability to allow an error comparable
in magnitude to the infidelity introduced by QPUs~\cite{Zhou2020,Pan2022,Ayral2022}. As a consequence, classical emulators of
quantum circuits based on tensor network methods serve as a bridge to quantum algorithms~\cite{qiskit,Villalonga2019,Nguyen2021,Ayral2022}.
In agreement with recent results discussing the aspect of entanglement in quantum
algorithms~\cite{Chen2022}, the amount of generated entanglement $-$ controllable via the
distribution of the entanglement entropy $-$ is another aspect to be
considered in addition to the finite fidelity: the two examples studied hereafter underline
that the green quantum advantage depends on both aspects.

In the following, we compare the energy consumption of different quantum computing architectures with a classical emulator. 
The classical numerical simulation methods considered are an evolution of the exact
state~\cite{JaschkeED} and tensor network (TN) methods~\cite{Orus2014,Montangero2018,Vidal2003,Schollwoeck2011,Nguyen2021, Haner2017}.
Tensor network methods are widely applied to a variety of quantum
many-body problems and benefit from their entanglement-based wave-function compression scheme. For pure quantum states, TN methods offer direct
access to the bipartite entanglement via singular value decomposition.
These features allow for an efficient simulation for states with a
limited amount of entanglement, e.g., for states following
the area law of entanglement~\cite{Eisert2010}.
Matrix product state (MPS) simulations for quantum circuits have convincingly
demonstrated the close connection between the entanglement generated by
quantum circuits and the ability to emulate the computation with an
MPS in the Google quantum advantage demonstration~\cite{Arute2019,Villalonga2020,Zhou2020}.
Beyond MPS methods obtaining the final state, tensor network methods relying
on an optimized contraction scheme have been introduced and successfully applied
to the question of quantum advantage~\cite{Pan2022}.

Our approach allows a direct prediction of the green quantum advantage threshold,
i.e., the \emph{Equality of Fidelity and Energy Point} (EFEP) between an MPS
and the quantum hardware, via a criterion
that can be calculated a priori as a function of the number of qubits, the generated
entanglement, and the QPU's gate properties to name the most important parameters.
In more detail, the final \efep{}s depend on various factors: the classical cost of a single two-qubit
gate application on an HPC system, the gate fidelities and
energy consumption of the quantum hardware, the circuit depth, and the number of
qubits, as well as the assumption of how much entanglement is generated within the
algorithm. 
Improvements in the efficiency per elementary classical operations have been pursued in the last decades~\cite{Hemmert2010}; and likewise, methods have been developed that reduce the energy for a pulse sequence driving an elementary quantum gate have been already developed~\cite{Ikonen2017}. Although these improvements are highly desirable, they do not change the scaling found hereafter which depends only on the gate fidelities and the entanglement generated during the quantum computation. 

Large-scale simulations are valuable to prove or disprove the quantum advantage for particular settings~\cite{Haner2017,Villalonga2020}, but every single simulation requires substantial HPC resources, and exploring a vast parameter landscape of high dimensionality is unfeasible. Moreover, small changes in the system parameters, such as gate fidelities, quickly add up to a drastic change in the final outcome, thus a fine-grained grid in the parameter landscape is desirable. Thus, here we develop an efficient way of mimicking the MPS simulation, i.e., an \emph{\mpsoracle{}}, which allows exploring parameter regimes otherwise beyond computational resources or only accessible with substantial computational costs and at a low resolution in the parameter space. 
We demonstrate the agreement of the exact MPS simulation with the \mpsoracle{}
in the relevant regimes for system sizes where the simulations are possible, i.e., the few-qubit case.
In conclusion, the developed methodology can be efficiently applied to
specific large-scale use-cases: we target NISQ
platforms and estimate the possible benefits for common quantum circuits,
e.g., the quantum Fourier transformation (QFT)~\cite{NielsenChuang}.

We estimate the classical computation's energy consumption in
Sec.~\ref{sec:class}. Then, we move on to the energy consumption of a
Rydberg simulator, an ion platform, and superconducting qubits
in Sec.~\ref{sec:quant}, before comparing the two
scenarios in different use-cases in Sec.~\ref{sec:usecase}. We present
our conclusion and outlook in Sec.~\ref{sec:concl}.

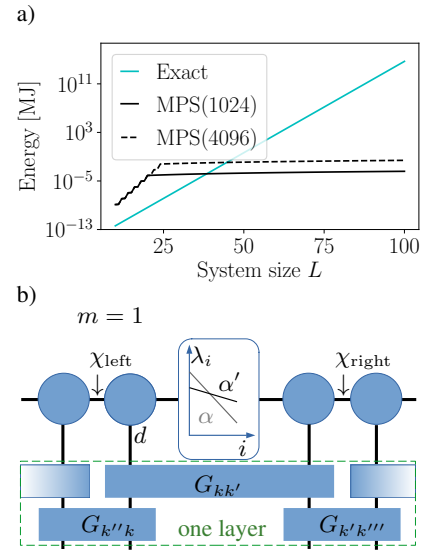
\begin{figure}[t]
  \begin{center}
    \vspace{0.8cm}\begin{minipage}{0.62\columnwidth}
      \begin{overpic}[width=1.0 \columnwidth,unit=1mm]{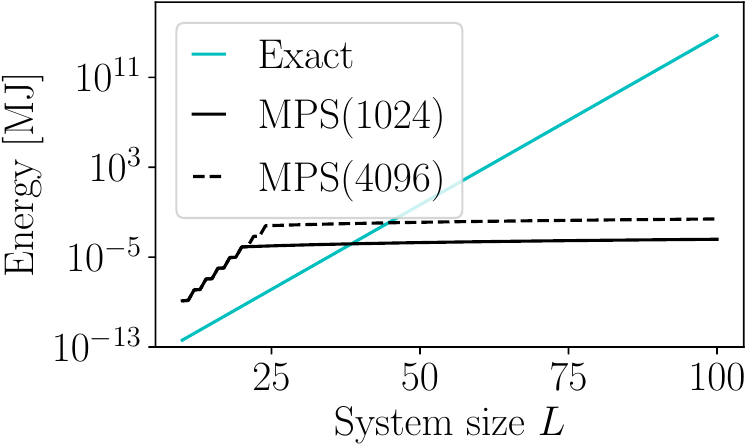}
        \put(0,  65){a)}
      \end{overpic}
    \end{minipage}
    \begin{minipage}{0.62\columnwidth}
      \vspace{0.6cm}
      \begin{overpic}[width=1.0 \columnwidth,unit=1mm]{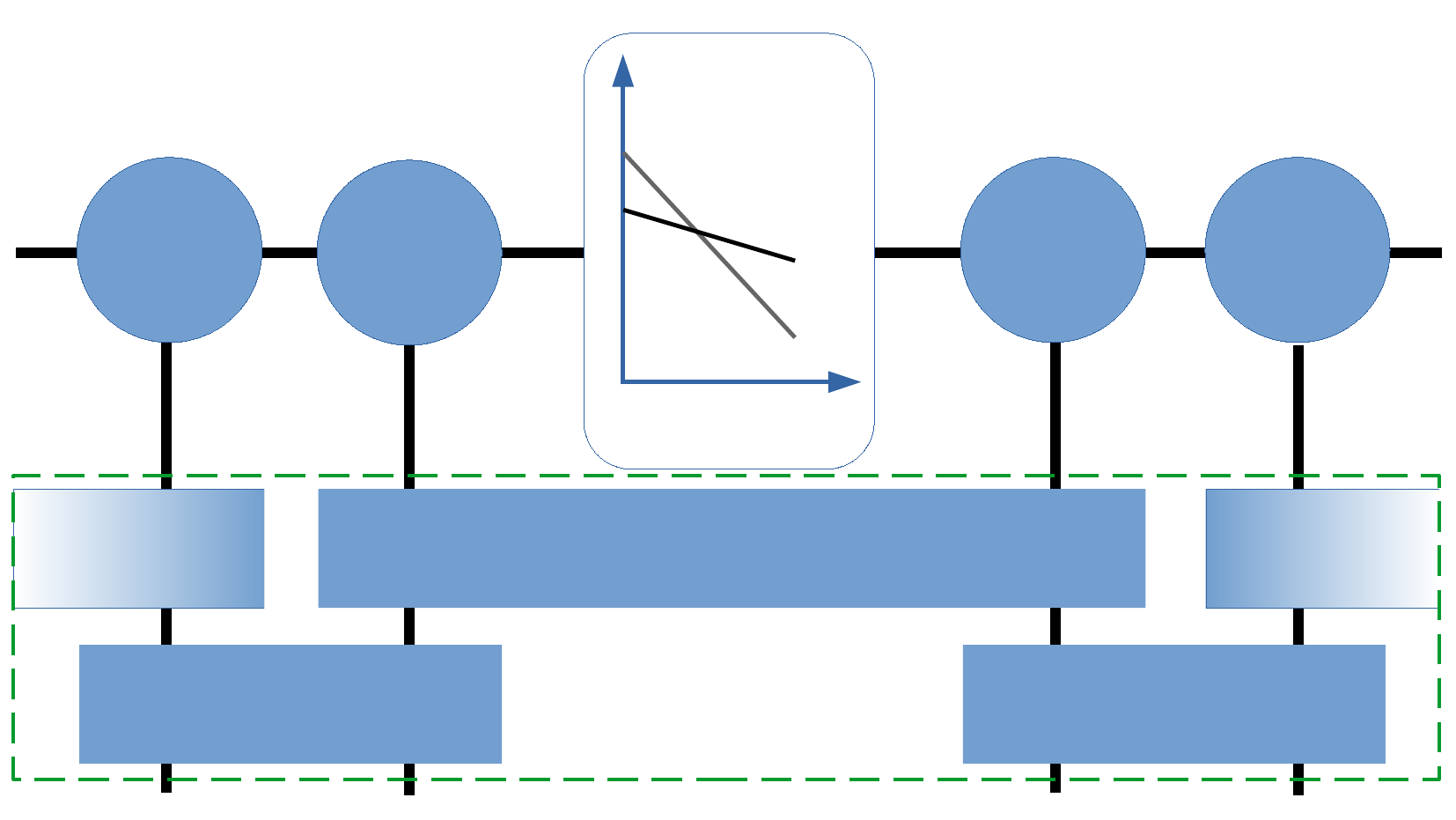}
        \put(0,  63){b)}
        \put(15,  58){$m = 1$}
        \put(45, 33){{\color{gray}$\alpha$}}
        \put(50, 41){$\alpha'$}
        \put(43, 48){$\lambda_{i}$}
        \put(55, 25){$i$}
        \put(29, 28){$d$}
        \put(18, 48){$\chi_{\mathrm{left}}$}
        \put(18.5, 41){$\downarrow$}
        \put(79, 48){$\chi_{\mathrm{right}}$}
        \put(79.5, 41){$\downarrow$}
        \put(40,  5){{\color{OliveGreen}one layer}}
        \put(16,  6){$G_{k''k}$}
        \put(75,  6){$G_{k'k'''}$}
        \put(44,  16){$G_{kk'}$}
      \end{overpic}
    \end{minipage}
    \caption{
      \emph{Tools for extracting the Equal Fidelity and Energy Points (EFEP).}
      a)~The starting point is the cost of applying
      a single two-qubit gate for the classical matrix product state emulator (MPS) for different bond
      dimensions $\chi$, see the label "MPS($\chi$)", and the exact state simulation, see label "Exact".
      b)~Instead of actually executing the classical MPS simulation, we
      refine this approach by sampling the values $\lambda_{i}$ of the Schmidt
      decomposition according to a postulated distribution, e.g., $\mathcal{D}_{p}(\alpha)$, after every two-qubit gate. This method allows us to tune the
      entanglement via different distributions, e.g., with a parameter $\alpha$. Then, we calculate an error based on the
      bond dimension $\chi$. The generic circuit with one layer, i.e., $m = 1$, acts on each pair
      of nearest neighbors once.
                                                                                \label{fig:classMJ}}
  \end{center}
\end{figure}

\section{Classical emulation of quantum computations                                                    \label{sec:class}}

The simulation of quantum computing systems via classical hardware can be grouped into different classes: exact state simulations and approximate methods, e.g., via tensor network methods exploiting the truncation of entanglement via the Schmidt decomposition~\cite{qiskit,Nguyen2021,Brennan2022,Ballarin2021}. These simulations are performed at the digital level, i.e., simulating directly 
quantum circuits as a sequence of quantum gates, or at the analog level of the computation, that is, solving the Schr{\"o}dinger equation via careful scheduling of the Hamiltonian terms and control fields driving the qubit dynamics in the hardware. An additional complexity level can be added considering the interaction with the environment, i.e., the full open quantum systems dynamics~\cite{BreuerPetruccione}. 
In this case, quantum gates turn into quantum channels and the
Schr{\"o}dinger equation into a master equation for open quantum systems, e.g., the Lindblad equation~\cite{Lindblad1976}. Here, we do not pursue the path of open quantum systems because the computational effort changes considerably and we point out further considerations in this scenario via either the purification of the system state or alternative strategies: the simulation of the Lindblad equation requires the same amount of resources as the simulation of the pure state with twice as many qubits. Alternatively, the simulations for the Lindblad equation are often performed via the quantum trajectories approach~\cite{Daley2014,Bonnes2014,JaschkeOTN}: an unraveling of the classical probabilities performed via an average over many simulations of pure state dynamics. In conclusion, hereafter we analyze the scaling of the energy cost of exact state and MPS-based digital simulations of pure states. 

\subsection{Computational cost for exact state evolution and matrix product states   \label{sec:comp_cost}}

We first estimate the energy cost of the exact state evolution of quantum circuits setting the reference for all other cases: as shown below, 
the number of floating-point operations is easily estimated 
and we extract the energy consumption of the most energy-efficient clusters from the Green500 list~\cite{Green2020}. The cluster efficiency has significantly improved over the past years: top-tier clusters consumed around $268 \MFLOPSW$ in 2012~\cite{Mammela2012}, improving from  $2.7 \MFLOPSW$ for standard clusters and $11.6 \MFLOPSW$ for energy-efficient designs in 2003~\cite{Feng2003}. In the following, we set today's HPC energy-efficient consumption to $20 \GFLOPSW$~\cite{Green2020}.
%
%

The wave function of $L$ qubits in a pure quantum state is represented
via $2^{L}$ floating-point (FLOP) numbers. The most straightforward application of a single-qubit gate, i.e., exploiting tensor notation, is equivalent to a matrix-matrix multiplication of dimension $2 \times 2$ and $2 \times 2^{L-1}$. These dimensions lead to $2^{L+2} \FLOP$, which already includes the required multiplication and summation. Similarly, a two-qubit
gate scales at $2^{L+3} \FLOP$, and a generic gate on $\eta$ qubits scale as $2^{L+\eta+1} \FLOP$. A FLOP refers here to complex-valued arithmetic. Thus, the application of a single one-qubit or two-qubit
gate costs approximately $7 \KJ$ or $14 \KJ$ ($1.9 \Wh$) for a system of 45 qubits, respectively. In general, we obtain for the $\eta$-qubit gate
%
%
\begin{align}
  E_{\mathrm{Gate}}^{\mathrm{HPC}}(\eta, L=45) &\approx 7 \cdot 2^{\eta - 1} \KJ \, .
\end{align}
This estimate covers the evolution via quantum gates but disregards the energy needed to compute measurements and non-necessary permutations~\cite{permutations}. Notice that the costs for communication, e.g., if using the Message Passing Interface (MPI), in first approximation are included in the energy consumption of the cluster per FLOP as the performance tests for HPC clusters are using parallel applications, e.g., a parallel version of the LINPACK benchmark~\cite{Dongarra2003}.

We now repeat the previous calculation for an approximate simulation performed by evolving an MPS.
The approximation introduced by the MPS approach consists of truncating the lowest singular values of any Schmidt decomposition which separates the system into a bipartition of two chains of qubits of length $L-M$ and $M$, respectively; the number of remaining singular values is the bond dimension $\chi$. The MPS simulation becomes exact once $\chi$ reaches $d^{\lfloor L/2 \rfloor}$, where $d$ is the Hilbert space dimension of a single site.
For each two-qubit gate, the MPS algorithm (i)~permutes sites
until they are nearest neighbors, (ii)~contracts neighboring sites, (iii)~contracts
the gate, and (iv)~splits the tensor related to the two involved qubits via a singular value decomposition (SVD).
Moreover, step~(i) contains the sub-steps a)~contract two sites, then
b)~split sites with swapped site indices via an SVD. There are $L - 1$
nearest-neighbors pairs, $L - 2$ next-nearest
neighbors pairs, until we reach one pair of sites being $(L-1)^{\mathrm{th}}$ neighbors. Assuming a random distance between the two qubits involved in the gates during a generic algorithm,  the average distance is
\begin{align}
  r(L)
  &= \frac{L + 1}{3} \, .
\end{align}
In conclusion, step (i) is composed on average by $F_{(i)}$ FLOP with
\begin{align}
  F_{(i)} &= (r(L) - 1) \left( 2 \chi^3 d^2 + 26 \chi^3 d^3 \right) \, ,
\end{align}
where the first term accounts for the contraction and the second for the SVD~\cite{Golub_VanLoan_96}; we recall that the system size is $L$, the dimension of the local Hilbert space $d$, and the bond dimension $\chi$. Moreover, permutations of the memory are neglected as they scale linearly for classical computers and, for simplicity, we assume that the bond dimensions $\chi$ are constant throughout the qubit chain. Similar estimates result in the following computational effort for the steps (ii) to (iv):
\begin{align}                                                                   \label{eq:F2_F4}
  F_{(ii)} &= 2 \chi^3 d^2 \, ,
  F_{(iii)} = 2 d^4 \chi^2 \, ,
  F_{(iv)} = 26 \chi^3 d^3 \, .
\end{align}
In conclusion, we obtain a total number of FLOP 
\begin{align}
  \FLOP_{\mathrm{\mathrm{Gate}}}^{\mathrm{Appr}} &= 26 r(L) \chi^3 d^3 + 2 r(L) \chi^3 d^2 + 2 \chi^2 d^4 \,.
\end{align}
The resulting average energy consumption per gate, under the assumption
of an energy consumption of $20 \GFLOPSW$, is reported in Fig.~\ref{fig:classMJ}a as a function of the number of qubits for a set of two different bond dimensions and compared with the exact state simulation. 
We
observe the bounded energy consumption of the MPS simulation in comparison to the exact state simulation: the FLOP count for the MPS increases in a controlled fashion as a function of the system size $L$ for a fixed
bond dimension $\chi$ and the computational resources are well below the exact state
simulation in the limit of large system sizes. When 
keeping only the terms with the highest-order scaling, the MPS approximately consumes
more energy than the exact simulation if
%
%
%
%
%
%
%
%
%
%
%
\begin{align}
  \chi &> \left( \frac{2^{L - 3}}{L} \right)^{\frac{1}{3}} \, .
\end{align}
The inequality yields a threshold for the bond dimension of $\chi \approx 1500$ for 40 qubits and confirms that lower-order terms play only an insignificant role.
This estimate is very coarse-grained and an upper bound for the MPS
effort~\cite{boundary_effects}, but nonetheless, the trends are valid including the fact
that an MPS becomes more expensive than the full state. We improve the
approach of estimating the effort through an \mpsoracle{} in the next subsection.

\subsection{Refined computational cost for matrix product states   \label{sec:comp_cost}}

In the following, we refine the FLOP and fidelity estimate via the introduction of the \mpsoracle{} -- without the need of running the full simulation -- exploiting one key property of an MPS simulation: the bond dimension grows dynamically over time and the bond dimension determines the numerical cost of the computation. 
Thus, to obtain an energy consumption estimate, we perform an \mpsoracle{} simulation that avoids evolving the state but - given a specific algorithm - tracks only the bond dimension of each link and the corresponding fidelity. The parameters of the \mpsoracle{} are the number of sites $L$, the dimension of the local Hilbert space $d$, the maximally allowed bond dimension $\chi_{\max}$, and a distribution $\mathcal{D}$ of singular values in the Schmidt decomposition $\lambda_{i}$. The sequence of gates is set by the algorithm of interest and the initial bond dimensions along the MPS are initially determined.

\begin{figure*}[t]
  \begin{center}
    \vspace{0.8cm}\begin{minipage}{1.95\columnwidth}
      \begin{overpic}[width=1.0 \columnwidth,unit=1mm]{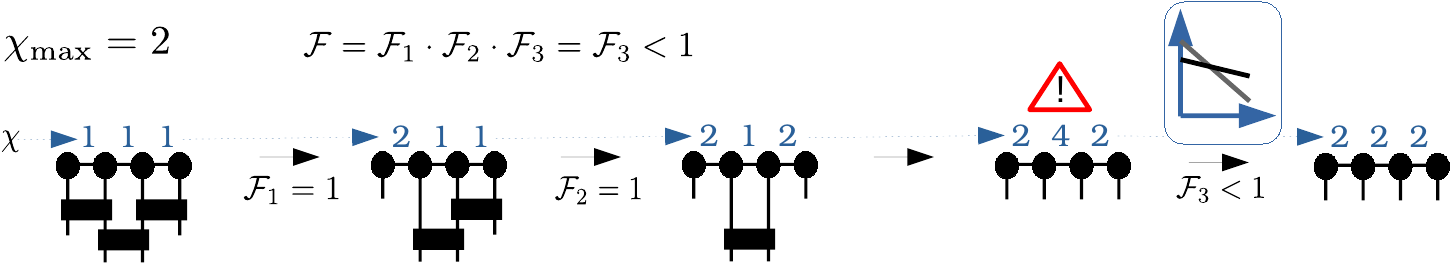}
        \put(83, 11.5){{\color{gray}$\alpha$}}
        \put(84, 14){$\alpha'$}
        \put(82, 16){$\lambda_{i}$}
        \put(87, 11){$i$}
      \end{overpic}
    \end{minipage}\hfill
    \caption{
      \emph{Four-qubit example for the mps-oracle.} While applying three
      two-qubit gates to an initial product state, one observes a growing bond
      dimension of the MPS. While the first two gates in steps 1 and 2 are applied
      without exceeding the maximal bond dimension, i.e., $\chi_{\max} = 2$
      in this example, the maximal bond dimension is exceeded after the third step and gate,
      as indicated by the warning sign. The
      mps-oracle samples from one of the available distributions $\mathcal{D}$ of singular values $\lambda_{i}$ as indicated
      by the sketch in the box 
      after the application of the third gate; this information is sufficient to estimate the truncation
      and reduce the bond dimension again to $\chi_{\max}$.
                                                                                \label{fig:mps_oracle_steps}}
  \end{center}
\end{figure*}
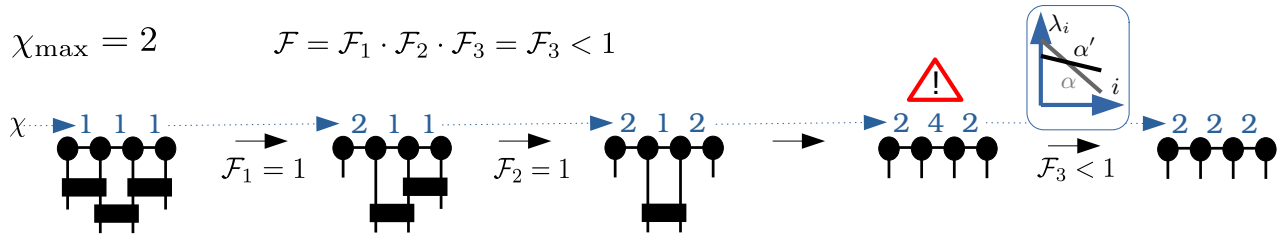

While iterating through the sequence of gates, we consider each gate
acting on two nearest neighbors and keep track of the decision if the MPS has to make a truncation based on the bond dimension of the previous step. Therefore, we compare the parameter $\chi_{\max}$ with the number of singular values yielded by the SVD, i.e., $\min(\chi_{\mathrm{left}} d, \chi_{\mathrm{right}} d)$.
If this minimum is larger than the parameter $\chi_{\max}$, we assume
that a truncation takes place according to singular values distributed
with $\mathcal{D}$ leading to a fidelity $\mathcal{F}_{k}$ for the $k$-th gate in the algorithm. 
Figures~\ref{fig:classMJ}b and \ref{fig:mps_oracle_steps} highlight these steps, the latter in a minimal example
of four qubits and three gates to be applied.
As shown in Fig.~\ref{fig:mps_oracle_steps}, no error is introduced when $\chi_{\max}$
is not reached, i.e., $\mathcal{F}_{k} = 1$. Combining the fidelities of each step $k$,
we obtain
\begin{align}                                                             \label{eq:cumulated_fidelity}
  \mathcal{F} = \prod_{k} \mathcal{F}_{k} \, ,
\end{align}
determining the fidelity of the complete algorithm via the
fidelity of its final state. Later, we use the
same equation to calculate the fidelity of an algorithm on each quantum hardware
and we plug in the fidelity of a single gate as $\mathcal{F}_{k}$.  This cumulative 
fidelity is a conservative estimate, which represents a worst-case scenario for
both MPS and QPU. We use
the infidelity $\mathcal{I} = (1 - \mathcal{F})$ to characterize the error.

Hereafter, we make the following assumptions about the algorithm and on the initial
state and explain a few restrictions which come with the \mpsoracle{}:\\
(i) We work with an uncompiled quantum circuit with
regard to the MPS connectivity; swaps are automatically executed by the \mpsoracle{} until the two sites of the two-qubit gates are nearest neighbors. Ordering of the two sites matters and is always installed via SWAPs in the \mpsoracle{}, i.e., we
neither swap the sites in the gate represented as a tensor nor do we keep track if gates are symmetric and swaps are superfluous. \\
(ii) The new bond dimension is obtained via the parameter $\chi_{\max}$ and
the neighboring bond dimensions $\chi_{\mathrm{left}}$ and $\chi_{\mathrm{right}}$.
Special properties of states cannot be tracked as there is no knowledge
of the state with our approach; for example, the application of a SWAP
gate to the GHZ state increases the entanglement predicted by the
\mpsoracle{}, despite $\mathrm{SWAP} \ket{\mathrm{GHZ}} = \ket{\mathrm{GHZ}}$
with equal entanglement before and after the application of the SWAP gate.
Another example would be the application of a gate and its inverse. Although
the overall entanglement in the system remains constant, the \mpsoracle{}
increases the entanglement estimate. \\
(iii) The
fidelity is computed per SVD as $\sum_{i=1}^{\chi_{\max}} \lambda_{i}^{2}$
for singular values $\lambda_{i}$ in descending order,
i.e., $\lambda_{i} \ge \lambda_{i+1}$. We propagate the fidelities
through the algorithm via Eq.~\eqref{eq:cumulated_fidelity} while assuming one of the following
unnormalized distributions for the singular values
\begin{align}
  \mathcal{D}_{e} &= \{ \mathrm{e}^{-i} \}, \, i = 1, \ldots, \chi \, ,          \label{eq:dist_exp} \\
  \mathcal{D}_{p}(\alpha) &= \{ i^{-\alpha} \}, \, i = 1, \ldots \chi \, ,       \label{eq:dist_pow} \\
  \mathcal{D}_{\mathrm{MP}} &= \{ \sqrt{\Lambda_{i}} \} , \, i = 1, \ldots \chi \, . \label{eq:dist_mp}
\end{align}
The distributions follow an exponential decay in Eq.~\eqref{eq:dist_exp},
a power-law behavior with the tunable parameter $\alpha$ as shown in
Eq.~\eqref{eq:dist_pow}, and the Marchenko-Pastur (MP) distribution for
the eigenvalues $\Lambda_{i}$ of random matrices \cite{Marchenko1967}, see
Eq.~\eqref{eq:dist_mp}. Figure~\ref{fig:classMJ}b represents
this step of choosing a distribution when splitting two neighboring sites
of an MPS into two: the index $i$ runs from $1$ to
$\chi = \min(\chi_{\mathrm{left}} d, \chi_{\mathrm{right}} d)$ and is cut
to $\chi_{\max}$ once the MPS exceeds $\chi_{\max}$ at any position.\\
(iv) As typical in most quantum algorithms, we assume in all the following examples that we start in a product state, i.e., $\chi =1 $.
Thus, the MPS emulator has a low computational cost, especially at the
beginning of the algorithm where the truncation error is small due to
starting in a product state.

This \mpsoracle{} can be used to estimate the energy consumption as a function of the target fidelity $\mathcal{F}$, while the bond dimension required for the EFEP is evaluated during this process. When used to evaluate the EFEP, the target fidelity is the value achieved by the currently available NISQ quantum hardware.

\section{Landscape analysis for quantum hardware                               \label{sec:quant}}

In the following paragraphs, we first explicitly describe our assumptions, e.g., on the measurement statistics, for the Rydberg platform. Then, we keep the same methods and assumptions used for the Rydberg hardware and analyze the trapped ions and superconducting platforms.

We neglect the different requirements in terms of compilation for each
platform, where the compilation converts the algorithm into a gate set
and connectivity suitable for each QPU. For example, a limited connectivity
introduces additional SWAP gates for two-qubit gates unavailable in the connectivity map.
This question becomes even more relevant once the hardware allows one to
modify the connectivity at run-time with shift operations.
We instead take into account the number of gates in a general formulation
of the algorithm and avoid this way the problem of finding the best available compiler
for each platform.

\emph{Rydberg quantum simulator $-$}
%
The estimate of the energy consumption on a Rydberg quantum simulator for a
complete algorithm $\erydalgo$ depends on the energy consumption of each
gate $\egate$, the number of repetitions $N$ required to obtain
the measurement statistics, and the number of gates $n$ for the algorithm
of interest:
\begin{align}                                                                   \label{eq:ealgo}
  \erydalgo &= n \cdot N \cdot \egate \, .
\end{align}
Herein, the number of gates is determined at a later stage after selecting
the algorithm. A first approximation of the number of repetitions is
$N = 1000$, which eventually depends on the fidelity and observable.
The energy consumed on a per-gate level requires knowledge of the power
consumption and the duration of each gate as explained next.

The consumption of each gate $\egate$ is estimated as
follows~\cite{Meinert2021,Pfau2021}: $10 \kW$ consumption by the experiment except
for the computers, which are taken into account with $0.5 \kW$
for approximately ten machines, i.e., in total $15 \kW$. Although the
pulse duration is around $100 \mathrm{ns}$, we consider the idle time in
between the gates; thus, the frequency at which we apply gates is
a better estimate, i.e., $\omega_{\mathrm{Gate}} = 1000 \mathrm{Hz}$.
In conclusion, a conservative estimate is
\begin{align}
  \erydalgo &\approx \frac{n \cdot N}{\omega_{\mathrm{Gate}}} \cdot 15 \kW
            = n \cdot 15 \kWs \, .
\end{align}
Thus, we obtain for the complete algorithm a fidelity
$\mathcal{F}_{\rm \scriptscriptstyle Algo}^{\rm \scriptscriptstyle Rydb} = 0.9988^{n}$
according to Eq.~\eqref{eq:cumulated_fidelity}, where $n$ is the number of gates.
The fidelity of an optimally controlled pulse
for a two-qubit gate, i.e., a theoretical value obtained through
numerical simulations, is $\mathcal{F}_{\rm \scriptscriptstyle Gate}^{\rm \scriptscriptstyle Rydb} = 0.9988$~\cite{Pagano2021,Pagano2022}. A difference
between one-qubit and two-qubit gates is not included; furthermore, we
assume the connectivity does not lower the fidelity.

We point out how variables scale to identify possibilities
to fine-tune our estimate for the Rydberg hardware. The laser energy
scales linearly with the number of qubits, i.e., more powerful lasers are
required to scale up the system~\cite{Henriet2020}. The selected Rabi
frequency changes the gate time, where the pulse becomes half as long if 
we double the Rabi frequency; increasing the Rabi frequency by a factor of two
induces a four-fold energy consumption of the laser. Thus, from a pure
viewpoint of the driving laser, the energy consumption doubles when reducing
the gate time from $\tau$ to $\tau / 2$. Although NISQ benefits from lower
gate times benefit, the Rabi frequency cannot be arbitrarily increased
without introducing other errors. These considerations can contribute
to a very fine-tuned estimate adapted for each specific hardware in the
future.

Before continuing with the following platforms, we explain that we neglect
certain contributions according to Tab.~\ref{tab:energy_per_step} which are not reflected in Eq.~\eqref{eq:ealgo}: the first step before
running an algorithm on the QPU is a compilation step. Compiling
requires classical computation and depending on the algorithm and
optimization steps of the compiler, the compilation consumes a
non-negligible computation time during which one consumes an energy
$\ecompile$. However, this step is an $\mathcal{O}(1)$ effort,
e.g., a QFT will be precompiled for any
platform. Nonetheless, this one-time effort can be considerable~\cite{Paler2022}.
This aspect changes if the algorithm needs to be recompiled
for every run due to changing parameters, e.g., for variational
quantum eigensolvers.
In the case of the Rydberg system, there is also an initialization step
loading the atoms into the optical tweezers; this loading step is necessary
for each run of the algorithm and consumes an energy $\eload$.
Another contribution neglected is the energy consumed in the
startup phase of the experiment, e.g., for cooling down the
apparatus or installing a vacuum. This step consumed an energy
$\estartup$ and has to be repeated after a time
$\tau_{\mathrm{uptime}}$, e.g., after maintenance. The fraction
of this energy $\estartup$ associated with each repetition of an
algorithm is then 
$\estartup \cdot \tau_{\mathrm{Algo}} / \tau_{\mathrm{uptime}}$,
where $\tau_{\mathrm{Algo}}$ contains all steps from loading the
atoms into the optical tweezers to the measurements. Comparing
the expected $\tau_{\mathrm{uptime}}$ of multiple weeks to the
duration of one algorithm, the contribution is insignificant.
The final step is the measurement: this contribution is not
considered either as the measurement follows the same logic as the
execution of gates; their power consumption times the
duration leads to a contribution $\emeas$ which can be added
to more detailed models. In addition, a more detailed model can tailor the number of
repetitions $N$ to obtain measurement statistics individually
for each algorithm. Thus, we suggest for future improved estimates,
but beyond the scope of this work, the following form of the
energy consumption for a Rydberg-QPU
\begin{align}
  \erydalgo &= N  \cdot \estartup \frac{\tau_{\mathrm{Algo}}}{\tau_{\mathrm{Uptime}}}
    + N \cdot n \cdot E_{\mathrm{Gate}} \nonumber \\
    &+ N \cdot E_{\mathrm{Measure}} + N \cdot E_{\mathrm{Load}} + E_{\mathrm{Compile}} \, .
\end{align}
\begin{table}[t]
  \begin{center}
  \caption{\emph{Energy consumption at different steps.} We consider the
    steps compile, startup, loading a trap, running the algorithm, and
    measurements for sampling $N$ times from the algorithm.
    We distinguish between negligible contributions of steps reused over
    a long time (\ding{216}), neglected contributions (\ding{55}),
    not applicable contributions ($-$), and the considered contributions
    of the actual circuit (\checkmark). Startup contains for example the process of creating
    a vacuum or cooling the superconducting hardware. For standard
    algorithms, a compiled instance of the circuit can be used multiple
    times; an equal argument applies to the contribution for startup.
    Loading the trap scales with the number of qubits, but not the number
    of gates.
                                                                                \label{tab:energy_per_step}}
  \begin{tabular}{@{} ccccc @{}}
    \toprule
    Step              & Repetitions      & Rydberg            & Ions               & Superconducting       \\
    \cmidrule(r){1-1} \cmidrule(rl){2-2}   \cmidrule(rl){3-3}   \cmidrule(rl){4-4}   \cmidrule(l){5-5}
    Compile           & $\le 1$          & \ding{216}         & \ding{216}         & \ding{216}            \\
    Startup           & $\ll 1$          & \ding{216}         & \ding{216}         & \ding{216}            \\
    Loading trap      & N                & \ding{55}          & \ding{55}          & $-$                   \\
    Run algorithm     & N                & \checkmark         & \checkmark         & \checkmark            \\
    Measurements      & N                & \ding{55}          & \ding{55}          & \ding{55}             \\
    \bottomrule
  \end{tabular}
  \end{center}
\end{table}

\emph{Trapped ion simulator $-$}
%
There are different trapped ion simulators available. For example, the setup
from \emph{Alpine Quantum Technologies} has the following
specifications~\cite{Pogorelov2021}: the gate fidelity of a single
M{\o}lmer-S{\o}rensen gate is $0.9983$ and has a pulse time of $200\mu{}s$;
the single-qubit Clifford gates reach a fidelity of $0.9983$; the
single-qubit rotations require a pulse time of $15\mu{}s$; the energy
consumption is bounded by $2 \cdot 3.7 kW$, which corresponds to the
two computing racks. We combine these numbers and insert them into
Eq.~\eqref{eq:ealgo} leading to the energy cost of this trapped ion platform:
\begin{align}
  \eaqtalgo &= N \left(0.015 n_1 + 0.2 n_2  \right) \frac{10^{-3}}{3600} \kWh \, .
\end{align}
We allow different parameters for single-qubit and two-qubit gates, where
$n_{1}$ and $n_{2}$ are the number of single-qubit and two-qubit gates,
respectively. The total number of gates of the algorithm is $n = n_{1} + n_{2}$.
The gate fidelities are comparable to the \emph{IonQ} platform; \emph{IonQ}'s
gate fidelities are on average around $99.5\%$ and $97.5\%$ in the one-qubit
and two-qubit gate, respectively~\cite{Wright2019}.

\emph{Superconducting quantum computers $-$}
%
The low energy consumption of Google's quantum computer has been already
discussed in \cite{Villalonga2020,Johansson2021}, which state 15kW and 26 kW
power consumption for the complete experiment. Reference \cite{Villalonga2020} also mentions that the
energy cost does not scale significantly with the number of qubits.

The pulse duration of the two-qubit gate is of the order of 12ns \cite{Arute2019}. One expects
gate fidelities of the order of $99.92\%$ and $99.4\%$ for single-qubit
and two-qubit gates, respectively~\cite{Barends2014}. We plug these numbers
into Eq.~\eqref{eq:ealgo} for the following analysis.
%
%
%
We expect similar energy consumption and performance for any other competitive
superconducting platform.


\section{Green quantum advantage                                                   \label{sec:usecase}}

We first analyze a generic algorithm, then focus on two different specific algorithms, the randomizing circuits used in Google's Sycamore
architecture~\cite{Arute2019} and the quantum Fourier transformation: we first analyze the entanglement generated on a small system and then study its scaling on larger system sizes via the \mpsoracle{}. 

\emph{Generic algorithm $-$}
%
We continue with the identification of the EFEPs. 
Hereafter, we tune the MPS-parameters via the \mpsoracle{} such that the fidelity and energy consumption of the classical emulator matches that of the
quantum hardware, i.e., as defined in the EFEP. Furthermore, we define one layer of the generic algorithm acting on $L$ qubits as $(L - 1)$ two-qubit gates acting on all nearest-neighbor in a 1-dimensional chain. If the gate $G_{k,k'}$ is acting on the qubits $k$
and $k'$, one layer of the algorithm for $L = 5$ is
$G_{1,2} G_{3,4} G_{2,3} G_{4,5}$, as sketched in Fig.~\ref{fig:classMJ}b for a bulk of a system.
To tune the circuit depth and number of gates, we apply $m$ of these layers;
thus, the total number of gates is $m (L - 1)$. With the odd-even and
even-odd pairs, we have a circuit depth of $2m$. We tune the entanglement
via $\alpha$ and the power-law distribution $\mathcal{D}_{p}(\alpha)$ in
Eq.~\eqref{eq:dist_pow}, where a lower (higher) value for $\alpha$ indicates
more (less) entanglement. This choice of a power-law decay is clearly the
less trivial and more relevant scenario, as the other two distributions,
i.e., exponential and MP behaviors, are clearly highly favorable for the
classical and quantum computation, respectively.

\begin{figure*}[t]
  \begin{minipage}{0.58\columnwidth}\raggedright a)
  \end{minipage}\hfill
  \begin{minipage}{0.58\columnwidth}\raggedright b)
  \end{minipage}\hfill
  \begin{minipage}{0.8\columnwidth}\raggedright c)
  \end{minipage}
  \begin{center}
    \begin{minipage}{0.58\columnwidth}
      \begin{overpic}[width=0.9 \columnwidth,unit=1mm]{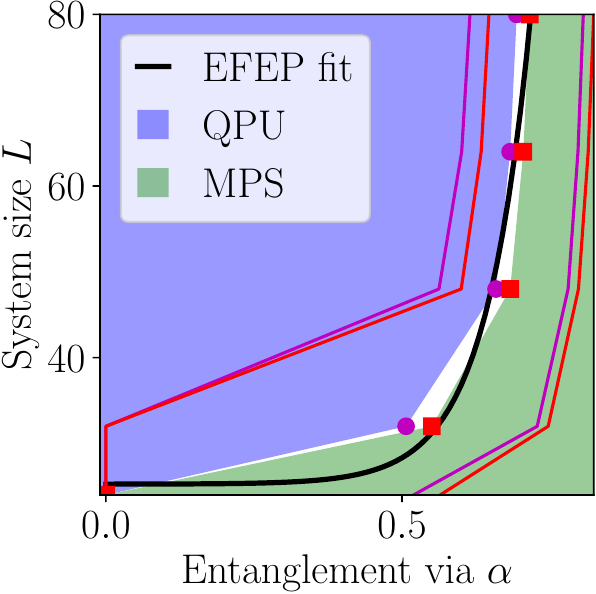}
      \end{overpic}
    \end{minipage}
    \begin{minipage}{0.58\columnwidth}
    \begin{overpic}[width=0.9 \columnwidth,unit=1mm]{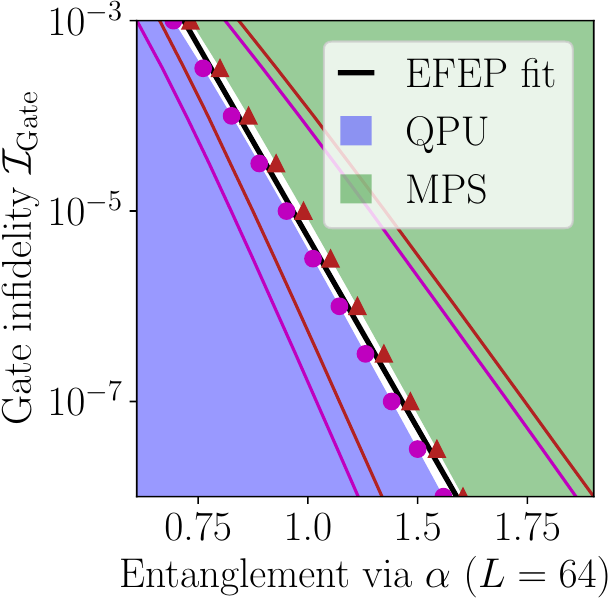}
    \end{overpic}
    \end{minipage}
    \begin{minipage}{0.8\columnwidth}
    \begin{overpic}[width=1.0 \columnwidth,unit=1mm]{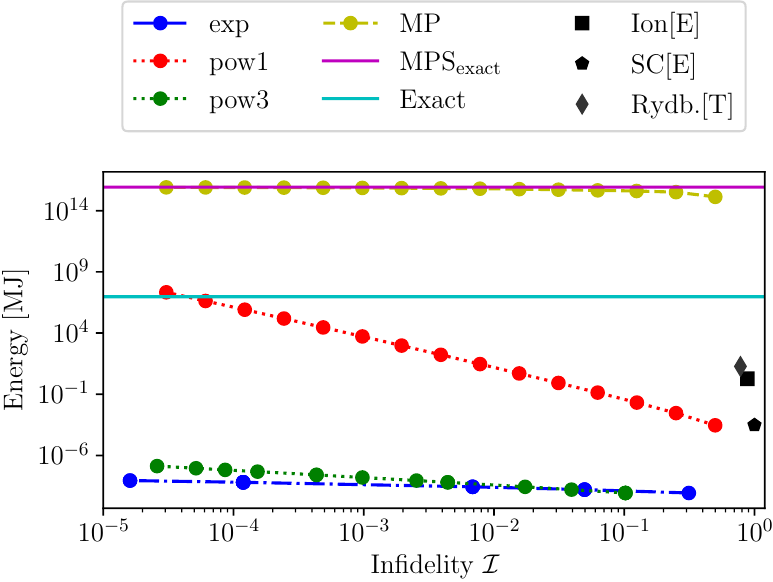}
    \end{overpic}
    \end{minipage}\hfill
    \caption{\emph{Generic algorithm regimes and distributions.}
      a)~We calculate the EFEPs for different power-law distributions $\mathcal{D}_{p}(\alpha)$ as a function
      of the system size $L$. From the viewpoint of the minimal energy consumption,
      the blue and green regions favor quantum processing units (QPUs) and MPS, respectively.
      The regimes do not shift significantly with the circuit depth as shown
      for $m$ layers. i.e., $m=12$ (purple circles) and $m=16$ (red squares). The curves of the same color as the data points indicate a factor of
      1000 difference in the energy consumption of quantum hardware and
      classical emulators. During the procedure, we vary the bond dimension $\chi$ and the entanglement
      via $\alpha$ to find the curve of EFEPs.
      Here, each of the $m$ layers
      acts first on the odd sites and their nearest neighbor and then
      on the even sites and their nearest neighbor.
      b)~We calculate the EFEPs for different power-law distributions $\alpha$ as a function
      of the gate infidelity. The QPU and MPS regimes are in the same color as for a) and
      we show the layers $m=12$ (purple circles) and $m=24$ (red $\blacktriangle$).
      c)~We add alternative distributions for the singular values in addition
      to the power law for $L = 64$ and conclude that the Marchenko-Pastur distribution (MP)
      is extremely expensive as expected, while exponential decays are favorable for classical
      emulators. Moreover, we compare the exact state (Exact), an MPS at full bond
      dimension, and the experimental platforms using trapped ions (Ion, experimental fidelities [E]),
      superconducting qubits (SC), or Rydberg atoms (Rydb, theoretical fidelities [T]).
                                                                                \label{fig:generic_regimes_fidelity}}
  \end{center}
\end{figure*}

Figure~\ref{fig:generic_regimes_fidelity}a indicates the two regimes for a fixed gate
fidelity of the Rydberg quantum hardware and different system sizes. The parameters
of the MPS emulator suggested by the \mpsoracle{}, i.e., the bond dimension $\chi$ and the decay
coefficient $\alpha$ for the singular values of Eq.~\eqref{eq:dist_pow},
lead to the EFEP. We easily
verify two expected trends: firstly, decreasing the system size $L$ extends
the regime of the MPS emulator. Secondly, increasing the number of layers $m$
favors quantum hardware, because the MPS has to operate for more steps
at a saturated bond dimension. The intriguing outcome of
Fig.~\ref{fig:generic_regimes_fidelity}a is the boundary around $\alpha = 0.75$ which
only changes insignificantly for large $L$. For clearness and to include the
necessary approximations and limits of our estimates with respect to the
experimental parameters and performances, we highlight also the boundary
where each approach consumes a factor of 1000 less energy than its counterpart:
as clearly shown, this factor only slightly shifts the boundary without
introducing a change in the scaling.
For large system sizes, a significant fraction is
still not accessible by the classical emulator, i.e., the area with
$\alpha < 0.6$. From this analysis, we propose $\alpha = 1$ as a safe
threshold to run algorithms on an MPS-emulator, e.g., within hybrid
quantum-classical algorithms. A fit of the EFEP boundary further supports this
hypothesis: we approximate the EFEP curve with
$L_{\mathrm{EFEP}}(\alpha) = 25.26 + 0.37 \cdot \exp(13.25 \cdot (\alpha - 0.34))$.
%
%
The fit is shown in Fig.~\ref{fig:generic_regimes_fidelity}a. An extrapolation yields a
range of $0.7 < \alpha < 1.0$ for system sizes reaching from $80$ to
$2000$ qubits.

\begin{table}[t]
  \begin{center}
  \caption{\emph{Entropy as a function of the \mpsoracle{} parameters.} We run the
  \mpsoracle{} for different system sizes $L$ to obtain the decay $\alpha$
  of the power-law distribution and the bond dimension $\chi$ at the EFEP for a
  generic algorithm. The resulting von-Neumann entropy $S$ of the
  subsystem for a bipartition of equal sizes serves as a guideline
  for the MPS vs. QPU decision for experiments where the entropy is
  available, but not the distribution of singular values. Analog to
  other results, the entropy shrinks for increasing system sizes and
  shifts the EFEP in favor of QPUs for larger $L$.
                                                                                \label{tab:entropy}}
  \begin{tabular}{@{} cccc @{}}
    \toprule
    System size $L$
    & Decay $\alpha$ 
    & Bond dimension $\chi$
    & Entropy $S$ \\
    \cmidrule(r){1-1} \cmidrule(rl){2-2} \cmidrule(rl){3-3} \cmidrule(l){4-4}
       $24$ & $6.3 \cdot 10^{-7}$ & $3935$  & $8.28$ \\
       $32$ & $0.55$             & $24021$ & $6.16$ \\
       $48$ & $0.68$             & $17137$ & $3.88$ \\
       $64$ & $0.70$             & $15925$ & $3.60$ \\
       $80$ & $0.72$             & $15177$ & $3.48$ \\
    \bottomrule
  \end{tabular}
  \end{center}
\end{table}

The entropy of a quantum system can be accessible to experiments even
when the distribution of singular values is out of reach for measurements~\cite{Notarnicola2021}.
Area laws for entanglement offer a common tool to describe and characterize
many-body quantum systems; the area law predicts a growing entropy
$S = -\mathrm{Tr} \left(\rho \, \mathrm{ln}(\rho) \right)$ for a subsystem's
density matrix $\rho$ as a function of the system size.
The EFEP does not follow an area law of entanglement:
Table~\ref{tab:entropy} contains data
for the EFEP of the generic algorithm for different system sizes, where the
power-law decay $\alpha$ and the bond dimension $\chi$ are chosen to match
the EFEP; the last column contains the entropy and is decreasing with system
size. Matrix product states emulators for larger system sizes compensate
for the advantage of the QPU with faster decaying power laws, i.e., $\alpha$
increases. As the bond dimension decreases at the same time, the entropy
reduces for larger systems at the EFEP. We point out that the data
for $L = 24$ in Tab.~\ref{tab:entropy} has no decay and therefore generates
more entropy even at a lower bond dimension. In summary, Tab.~\ref{tab:entropy}
serves as a rule-of-thumb of the EFEP if only the entropy is available.

For the next generations of quantum hardware, we are interested in
how much an improvement of the fidelity shifts the border in
favor of quantum hardware, see Fig.~\ref{fig:generic_regimes_fidelity}b.
Starting at a single gate fidelity of approximately $10^{-3}$ and
the boundary at $\alpha \approx 0.7$, the boundary shifts towards
$\alpha \approx 1.3$ for a single gate fidelity around $10^{-7}$ for
a system size of $L = 64$ at a fixed number of layers, i.e., $m=24$.
This shift is significant, e.g., we have the two following sets of
parameters at the EFEP with
$\mathcal{I}_{\mathrm{Gate}} = 10^{-3}, \alpha = 0.73, \chi = 14714, S = 3.29$
and
$\mathcal{I}_{\mathrm{Gate}} = 10^{-6}, \alpha = 1.35, \chi= 14625, S = 0.85$.
We observe that the bond dimension stays almost equal as the energy
consumption of the hardware does not change in this scenario. As the
fidelity of the quantum hardware improves, the MPS has to capture the
entropy $S$ more precisely, which leads to a faster-decaying distribution
of singular values and, thus, we remain with much less entropy.
A numerical fit delivers an estimate of the infidelity of a single
gate $\mathcal{I}_{\mathrm{Gate}}^{\mathrm{EFEP}} \approx 10^{(2.77 - 8.04 \alpha)}$.


The generic algorithm provides a lot of degrees of freedom in terms of tuning
the generated entanglement, the system size, and the gate fidelity on a QPU.
In contrast, Fig.~\ref{fig:generic_regimes_fidelity}c fixes one specific system
size of $L = 64$ to compare the three different quantum platforms to
the exact state, the exact MPS at full bond dimension, and the different
distributions via the \mpsoracle{}. The data underlines our previous statement
that the Marchenko-Pastur distribution is extremely costly on a classical
emulator independent of the target fidelity. In the other extreme, exponential
and fast decaying power-law distributions are favorable for an MPS scenario.
The difference in the energy costs of each hardware platform originates
in the time to apply a single gate, which differs by almost five orders of
magnitudes between the platforms. In contrast, the power consumption
is within a factor of three. The infidelity of the QPU platforms has to
considered with regard to the conservative estimate of the fidelity of the
final state according to Eq.~\eqref{eq:cumulated_fidelity}; quantum error
correcting codes that require a ten-times or a hundred-times overhead in the number
of gates will change the fidelity significantly while an additional two orders of
magnitude for the quantum error correcting operations will not change the energy
on the $y$-scale significantly.

\emph{Randomizing circuits $-$}
%
%
\begin{figure*}[t]
  \begin{minipage}{0.24\linewidth}\raggedright a)
  \end{minipage}\hfill
  \begin{minipage}{0.24\linewidth}\raggedright b)
  \end{minipage}\hfill
  \begin{minipage}{0.47\linewidth}\raggedright c)
  \end{minipage}
  \begin{center}
    \begin{minipage}{0.24\linewidth}
      \begin{overpic}[width=1.0 \columnwidth,unit=1mm]{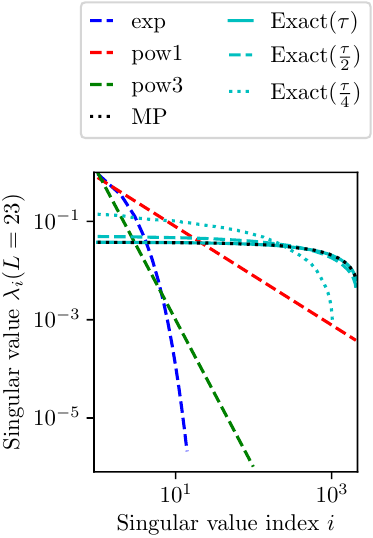}
      \end{overpic}
    \end{minipage}\hfill
    \begin{minipage}{0.24\linewidth}
      \begin{overpic}[width=1.0 \columnwidth,unit=1mm]{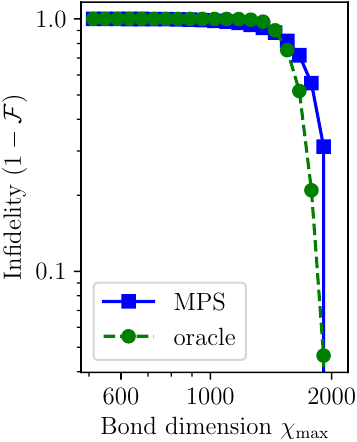}
      \end{overpic}
    \end{minipage}\hfill
    \begin{minipage}{0.47\linewidth}
      \begin{overpic}[width=1.0 \columnwidth,unit=1mm]{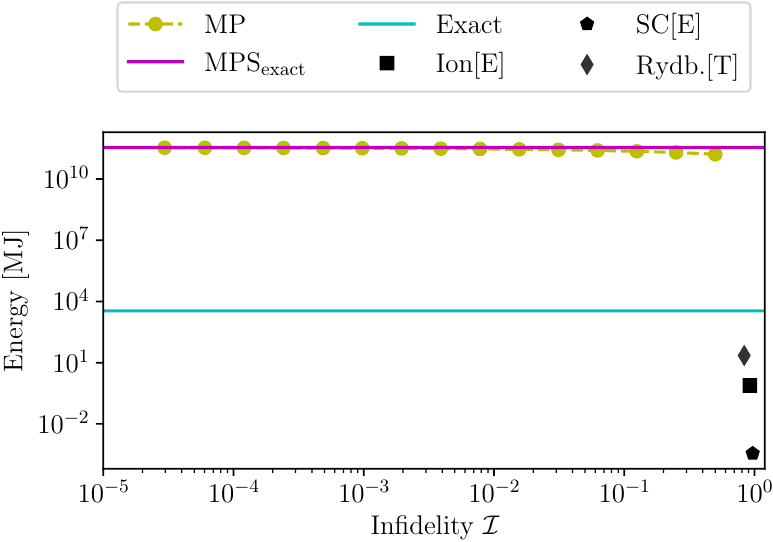}
      \end{overpic}
    \end{minipage}
    \caption{\emph{Random gates circuit.} We analyze the sequence
      of random gates used for Google's Sycamore platform. The number
      of layers is $\mrnd = 20$.
      a)~The singular values for a reduced version of the Sycamore
      architecture with $23$ qubits indicate which distribution of
      singular values we should expect for this algorithm: the
      Marchenko-Pastur (MP) distribution is the closest for the
      given options. The exponential decay (exp) and the $1/r^3$ power law (pow3)
      are far away from the true singular values. A $1/r$ power
      law (pow1) still underestimates the singular values. In addition
      to the final distribution of the exact state simulation at time $\tau$ (Exact($\tau$)), we show
      the distribution after a quarter and a half of the evolution.
      b)~For low bond dimensions, we compare the estimate of the
      \mpsoracle{} (oracle) with regard to the fidelity with the actual MPS simulation.
      The estimate of the \mpsoracle{} yields a reasonable estimate.
      The large truncation error even for high bond dimensions is related to the
      Marchenko-Pastur distribution for the singular values.
      c) The energy consumption of random gates used for Google's Sycamore platform
      in $\MJ$ as a function of the infidelity yields a clear separation in energy.
      The choice of the MP distribution for the singular values in a
      matrix product state has a significant effect on the FLOP count;
      here, we observe that the MPS emulator becomes unfavorable due
      the high entanglement within this approach.
      The reference points are the Rydberg system (Rydb.) with the theoretical ([T])
      fidelity from the optimal control simulations, the upper bound of
      energy consumption for trapped ions (Ion) with experimental ([E])
      gate fidelities, the superconducting qubits (SC), and the complete
      state vector, see the label "Exact".
                                                                                \label{fig:prestudy}}
  \end{center}
\end{figure*}
We now examine the randomizing circuits used in
Google's Sycamore architecture~\cite{Arute2019}. The
quantum circuit acting on 53 contains a single-qubit gate on each qubit
followed by 20 to 24 two-qubit gates, which are selected randomly out of
four subgroups with compatible connectivity. The complete sequence
contains $\mrnd$ repetitions of these gates. The same sequence is also used
in the experiment in~\cite{Wu2021}. For the \mpsoracle{}, the algorithm
is not optimized beforehand via a transpiler for the connectivity of a
one-dimensional chain; the MPS sites are ordered in a zig-zag scheme. The zig-zag
scheme maps a two-dimensional grid to a one-dimensional system by walking along each row of the
matrix always starting at the first column and moving towards the
last column.
With a fixed algorithm at hand, we first answer
the question of which distribution of singular values we
choose for the randomizing circuits. We compare the exact simulation with the
distributions from Eqs.~\eqref{eq:dist_exp} and \eqref{eq:dist_mp} with
twenty layers of the random gates in Fig.~\ref{fig:prestudy}a; we reduce
the number of qubits to $L = 23$ to perform fast exact state simulations and compare
the actual singular values at time $\tau$, i.e., the end of the evolution.
According to the data at the end of the gate sequence, we have to consider
a Marchenko-Pastur distribution for the singular values. Additional
data points at $\tau / 2$ and $\tau / 4$ demonstrate how the entanglement
grows over time and remains below the Marchenko-Pastur distribution
at the intermediate points.
Figure~\ref{fig:prestudy}b compares the fidelity calculated from the
final states of an actual MPS simulation~\cite{Ballarin2021} between
$\ket{\psi(\chi_{\mathrm{full}})}$ and $\ket{\psi_{\mathrm{MPS}}(\chi)}$ with
the fidelity predicted by the \mpsoracle{}. We define the bond dimension
$\chi_{\mathrm{full}} = d^{\lfloor L/2 \rfloor}$. The data indicates 
two regimes, i.e., the first regime where a high truncation due to
the MP distribution leads to a final fidelity close to $1.0$
despite the fact of covering a considerable fraction of $\chi_{\mathrm{full}}$.
The second regime of low infidelities is especially of interest for future
hardware with increasing fidelities and the \mpsoracle{} provides us
with a reasonable prediction.
A more sophisticated error model is beyond the scope of this work.

Figure~\ref{fig:prestudy}c captures the effects of the
MP distributions for the singular values with $L = 53$, which needs a
bond dimension almost at saturation of $d^{\lfloor L/2 \rfloor}$.
Recall that the estimate for the Rydberg hardware uses theoretical predictions
for the gate fidelity. We emphasize that the ion
data points and the results for the superconducting qubits refer to experimental
fidelities and therefore cannot be compared to the Rydberg platform on equal
footing. On the Rydberg hardware, experimental gate fidelities of $97.4\%$
have been reached\cite{Levine2019}.
In summary, the data underlines why this experiment has been considered
to prove quantum advantage due to the amount of entanglement present in
the system.

\emph{Quantum Fourier transformation $-$}
%
Now, we turn to the second example and investigate the QFT algorithm. We
consider $\mqft$ repetitions of the algorithm which allows us to tune the
generated entanglement while starting in a product state. A single
QFT executed on a product state does not generate enough entanglement
for a meaningful comparison; we can tune the number of QFTs more easily
than the entanglement in the initial state. In the following, we set the
number of layers to $\mqft = 4$, where each layer contains the
complete QFT algorithm.

\begin{figure}[t]
  \begin{center}
    \vspace{0.5cm}\begin{minipage}{0.77\columnwidth}
      \begin{overpic}[width=1.0 \columnwidth,unit=1mm]{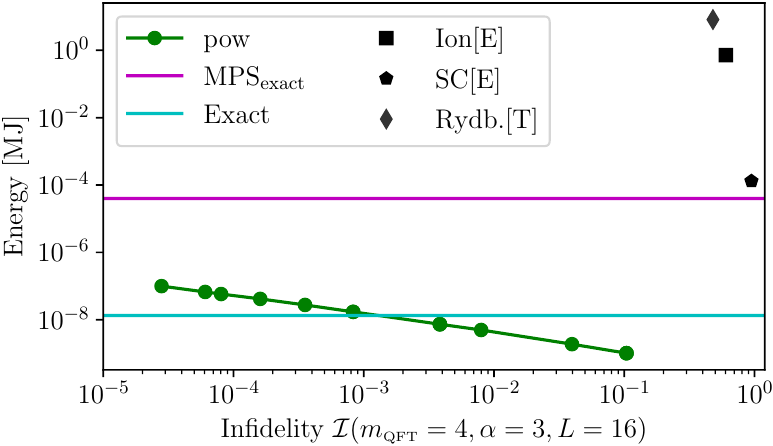}
        \put(0,  60){a)}
      \end{overpic}
    \end{minipage}\hfill
    \begin{minipage}{0.77\columnwidth}\vspace{0.5cm}
      \begin{overpic}[width=1.0 \columnwidth,unit=1mm]{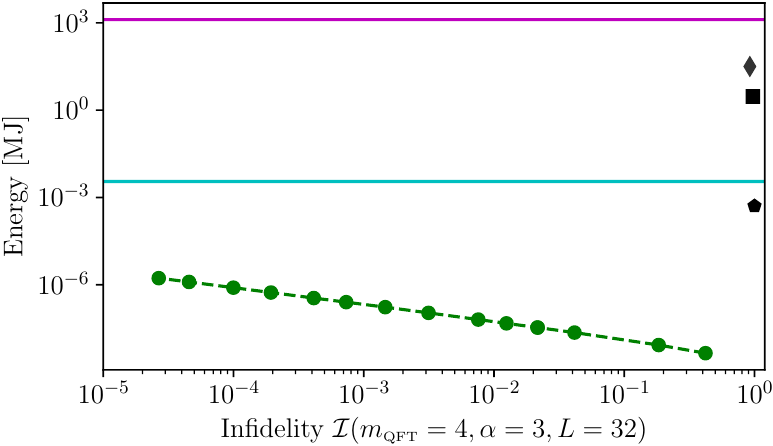}
        \put(0,  60){b)}
      \end{overpic}
    \end{minipage}\hfill
    \begin{minipage}{0.77\columnwidth}\vspace{0.5cm}
      \begin{overpic}[width=1.0 \columnwidth,unit=1mm]{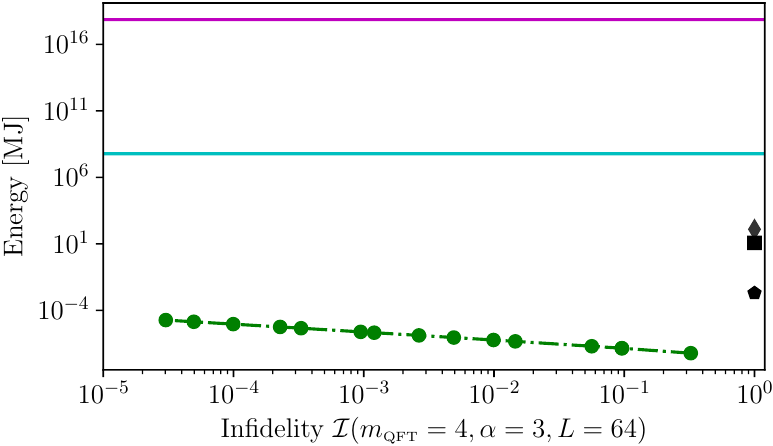}
        \put(0,  60){c)}
      \end{overpic}
    \end{minipage}
    \caption{\emph{Quantum Fourier transformation.} We apply multiple
      iterations of the quantum Fourier transformation and compare
      the predicted energy consumption by classical emulators and
      the Rydberg platform (Rydb., theoretical gate fidelity),
      trapped ions (Ion, experimental gate fidelity [E]), and
      superconducting qubits (SC). We consider three different system sizes:
      a)~For $L=16$ sites, the emulator
      via MPS or the exact states is favorable in comparison to the
      quantum hardware, where the singular values decay as a power law
      for the QFT.
      b)~For $L=32$, the quantum hardware becomes more
      energy-efficient than the complete state.
      c)~For $L=64$, the range of energies is now beyond twenty orders
      of magnitude. The relative difference between the \mpsoracle{} with a
      power-law distribution $\mathcal{D}_{p}(\alpha=3)$ and the
      quantum hardware decreases slightly.
                                                                                \label{fig:qft}}
  \end{center}
\end{figure}

We evaluate in Fig.~\ref{fig:qft_sv_distr} in App.~\ref{sec:sv_distr} which distribution
from Eqs.~\eqref{eq:dist_exp} to \eqref{eq:dist_mp} represents best the QFT
algorithm and choose the power-law distribution. Unlike the randomizing circuit and
its MP distribution, the power-law distribution does not
match the singular values from the QFT as well.
The power-law distribution with a coefficient of
$\alpha = 3$ leads to a regime where the MPS is more competitive than for
the Marchenko-Pastur distribution, i.e., we expect bond dimensions $\chi$
below $d^{\lfloor L/2 \rfloor}$.

We compare across different system sizes from $L = 16$
to $L = 64$ in the example of the QFT; the comparisons between the \mpsoracle{} and quantum hardware are shown
in Fig.~\ref{fig:qft}. We plot the energy consumption of
the tensor network emulator as a function of the fidelity. 
Of interest is the order of magnitude in terms of the energy consumption, e.g., the vertical
difference between the Rydberg platform data points and the data points
for the \mpsoracle{} using the power-law decay. On the horizontal axis, the
improvement in fidelity for the complete sequence
of gates is the horizontal difference to the closest data point of the
\mpsoracle{} prediction.
For comparison, we show the Rydberg reference points when using a
theoretical ([T]) gate fidelity of
$\mathcal{F}_{\rm \scriptscriptstyle Gate}^{\rm \scriptscriptstyle Rydb} = 0.9988$,
the trapped ion upper bound with experimental ([E]) fidelity, the superconducting
qubits with experimental fidelities, and the exact state
vector. The exact state vector has by default an infidelity
of zero and is therefore visualized as a constant line. The second point of
reference and constant line with fidelity $\mathcal{F} = 1$ is the MPS at a bond dimension where the MPS becomes exact.

We recognize different regimes in the different system sizes. As
expected, small systems favor classical emulators and Fig.~\ref{fig:qft}a
for $L = 16$ represents the first regime, where the quantum hardware
has a higher predicted energy consumption in comparison to any emulated simulation.
In fact, even the worst-case scenario of an MPS at full bond dimension
is slightly better than the quantum platforms.
The method of choice is either the exact state or
the MPS depending on the target fidelity of the full algorithm.
For $L = 32$, the situation changes and
the quantum hardware is from an energy standpoint distributed around the
exact state method depending on the platform. The scenario with $L = 64$
changes the situation: both emulator and quantum hardware
are orders of magnitudes cheaper than the exact state method. The ration
of $E_{\rm \scriptscriptstyle QPU} / E_{\rm \scriptscriptstyle MPS}$ slightly
decreases, which is not visible in Fig.~\ref{fig:qft}
covering several orders of magnitude in energy.

Moreover, we predict with Fig.~\ref{fig:qft}c
that the Rydberg platform amongst others has to improve by several orders of magnitude
with regard to the energy consumption and one order of magnitude in
the gate fidelity to outperform the tensor
network simulation for the QFT algorithm. The singular values in the simulation follow a
power law with a cubic decay, which is favorable to tensor network simulations.

As a side remark for the QFT, we point out that the algorithm is rarely run
on its own, e.g. the QFT is used within Shor's algorithm. If executed after
a highly entangling part of the circuit, the computational scaling of the
QFT also increases.

\section{Conclusion                                                            \label{sec:concl}}

We have compared the energy consumption of quantum hardware
with a classical emulator of the quantum algorithm. We address the
problem from the viewpoint of reaching the \emph{Equal Fidelity and Energy Point} (EFEP) of the two platforms.
Thus, the fidelity of a single gate is crucial for both, the quantum computer
and the MPS emulator. Low gate
fidelities accumulate to a large error; the
distribution of singular values is the key parameter on any MPS emulator.
The \mpsoracle{} enables us to predict the computational effort of
quantum circuits emulators running on HPC or as a decision-making rule
for possible applications in hybrid quantum-classical algorithms and
architectures.

The randomizing circuit generates an
amount of entanglement which makes the quantum hardware favorable;
the main reason for this characteristic is the singular value distribution of the Schmidt values according to the
Marchenko-Pastur distribution. In contrast, if
the singular values follow a power-law distribution, tensor network simulations
via matrix product states are competitive for values of $\alpha$
greater than $0.75$ to $1.0$ depending on the fidelity of a single
gate. The latter case of a power-law distribution appears for example
in an isolated quantum Fourier transformation. We provide fits for
the EFEP curve, which underline the aforementioned range, where,
in our example, the value of $\alpha = 1.0$ is reached at around
$2000$ qubits according to the extrapolation. The fit of the EFEP curve
as a function of the single-gate fidelity predicts an $\alpha \approx 1.5$
once the fidelity for a single-gate reaches a $10^{-7}$ level.

The presented methodology evaluates the relative
energy consumption of tensor network emulators and
quantum hardware. The following open questions lead to a more detailed picture
while adding more specific details: 
we stress that quantum computers can also improve their energy consumption
by performing the gate sequence with the
same fidelity and power consumption in less time, e.g., by performing
gates in parallel or combining gates leading to a lower pulse
time~\cite{Gu2021};
one can go beyond
the fidelity point of view and compare the measurement statistics
of classical and quantum simulations. The application to adiabatic quantum computing represents
an additional direction, but the error calculations on the classical
simulation side and the quench on the quantum hardware have to be taken
into account for the \mpsoracle{}.
Another question is whether open systems should be included
in the considerations, which increase the computational cost of the classical
simulation; in the scenario of a classical emulator, the absence of
decoherence might be favorable. Finally, one can move to a more and more
detailed picture where, in the end, our work and its perspective on the
energy consumption might directly influence the future development 
of hybrid quantum-classical applications.

\section*{Acknowledgments}

We gratefully appreciate the contributions from Florian Meinert
and Tilman Pfau on the energy consumption of the Rydberg platform. We
thank Marco Ballarin, Phila Rembold, and Pietro Silvi for useful discussions and feedback.

We acknowledge funding from the German Federal Ministry of
Education and Research (BMBF) under the funding program quantum
technologies $-$ from basic research to market $-$ with the grant QRydDemo, the Italian PRIN 2017, the EU-QuantERA projects QuantHEP and T-NISQ, the EU project TEXTAROSSA, and the WCRI-Quantum Computing and Simulation Center of Padova University.


\begin{thebibliography}{60}
\bibitem [{per()}]{permutations}%
  \BibitemOpen
  \href@noop {} {\enquote {\bibinfo {title} {{The application of gates changes
  the order of the qubits. We refer to the non-necessary permutations as a
  permutation that installs the original order of qubits after the application
  of a gate. Instead, it is cheaper to keep track of each qubit's position
  taking into account all the permutations induced by the gates applied so
  far.}}}\ }\BibitemShut {NoStop}%
\bibitem [{bou()}]{boundary_effects}%
  \BibitemOpen
  \href@noop {} {\enquote {\bibinfo {title} {{The estimate is an upper bound
  for the MPS since the approach does not take into account any boundary
  effects and, thus, our estimates are unfavorable for the MPS, e.g., a fixed
  bond dimension $\chi$ along the system. The fixed bond dimension along the
  MPS is an unfavorable assumption because the actual maximal bond dimension to
  capture an exact state depends on the number of left neighbors $n_l$ and
  right neighbors $n_r$ for each link and is then $d^{\min(n_l, n_r)}$, i.e.,
  exponential growth towards the center of the MPS. The argument that the MPS
  can become more expensive is based on the application of a gate at the middle
  of the MPS. The construction in step (ii) requires $2 d^{L/2 - 1} d d^{L/2} d
  d^{L/2 - 1} = 2 d^{3L / 2}$ FLOP, where we resolved the $\chi^3$ from
  Eq.~\eqref{eq:F2_F4} with the actual variable bond dimension along the MPS.
  In contrast, the exact state requires $2 d^{L + 2}$ FLOP. Thus, alone the
  step (ii) is more costly for an MPS representing the full entanglement than
  the step on the complete state vector for $L > 4$.}}}\ }\BibitemShut
  {NoStop}%
\end{thebibliography}
%

\appendix

\section{Marchenko-Pastur distribution for the Schmidt decomposition           \label{sec:marchenko}}

We use the Marchenko-Pastur distribution $\mathcal{D}_{\mathrm{MP}}$ \cite{Marchenko1967} to generate
singular values according to random quantum states because the distribution is the most
precise way to generate samples of singular values in comparison to the other
approaches in the case of the Sycamore random gate sequence.
As the Marchenko-Pastur distribution describes the eigenvalue distribution
of a matrix in which the entries are distributed randomly according to a Gaussian
distribution with zero mean, we have to make additional adaptions. We start with
real-valued wave functions and generate a vector $r$ of random numbers with
$u \in \mathbb{R}^{d_{1} \cdot d_{2}} \sim \mathcal{N}(0, \sigma^2)$. We choose
$\sigma = 1 / \sqrt{2^{d_{1} \cdot d_{2}}}$, which leads to a normalized wave
function with $u^{\mathrm{T}} u = 1$ in the limit of $d_{1}, d_{2} \to \infty$.
To ensure normalization, we actually define
$\ket{\psi} = u / \sqrt{u^{\mathrm{T}} u}$. We follow the standard procedure
to obtain the Schmidt decomposition of the quantum state $\ket{\psi}$ for
the two subsystems of dimension $d_{1}$ and $d_{2}$ and define the
corresponding matrix $M \in \mathbb{R}^{d_{1} \times d_{2}}$ for the singular
value decomposition. The SVD yields $\min(d_{1}, d_{2})$ singular values
$\lambda_{k}$; the values $\lambda_{k}^{2}$ are distributed according to
the Marchenko-Pastur distribution. The original statement is in terms
of the eigenvalues $\Lambda_{k}$ of the matrix $M M^{\mathrm{T}}$; it holds
$\Lambda_{k} = \lambda_{k}^{2}$ using the unitary characteristics of the
SVD.

\begin{figure}[t]
  \begin{center}
    \vspace{0.8cm}
    \begin{overpic}[width=1.0 \columnwidth,unit=1mm]{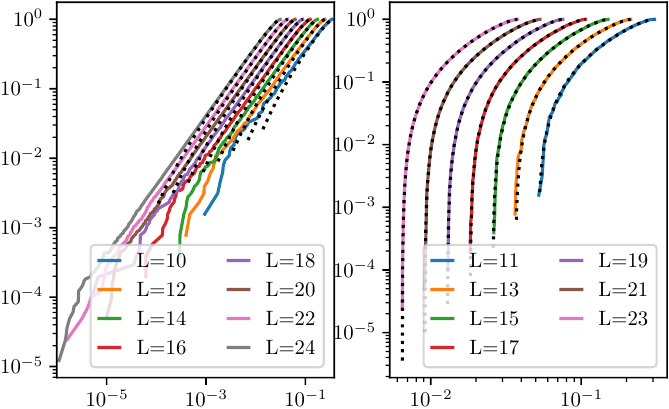}
      \put( 0,65){a)}
      \put(50,65){b)}
    \end{overpic}
    \caption{\emph{Marchenko-Pastur distribution.} We compare the average
      cumulative probabilities for randomly generated quantum states to the
      cumulative probabilities of the Marchenko-Pastur distribution for the
      empirically chosen parameter $\sigma$, which are shown as black dashed
      lines. Due to their different shapes, we show even system sizes in
      a) and odd system sizes in b). Note that the deviations in a) are in
      a range of $10^{-5}$ to $10^{-2}$- for the $x$-axis, which are not
      present for the odd system sizes in b).
                                                                                \label{fig:marchenko}}
  \end{center}
\end{figure}

The real-valued wave function restricts the generality of quantum mechanics.
Therefore, we move towards complex-valued wave functions. We generate two
random vectors
$u_{1}, u_{2} \in \mathbb{R}^{d_{1} \cdot d_{2}} \sim \mathcal{N}(0, \sigma^2)$
and combine them to a complex-valued random vector as
$v = u_{1} + \mathrm{i} u_{2}$. The wave function is defined analog to
the real-valued case with $\ket{\psi} = v / \sqrt{v^{\mathrm{T}} v}$.
We empirically find the standard deviation
$\sigma$ to be used in the Marchenko-Pastur distribution for a Schmidt
decomposition in the center of the system with
\begin{align}
  \sigma_{\mathrm{even}} &= \sqrt{2^{-L}} = \frac{1}{d_{1}} \, , \qquad
  \sigma_{\mathrm{odd}} &= \max \left( \frac{1}{d_{1}}, \frac{1}{d_{2}} \right) \, .
\end{align}
The validation for different system sizes is shown in Fig.~\ref{fig:marchenko},
where we consider the Schmidt decomposition at the center of the system
for system sizes $L = 10, \ldots, 24$ averaged over 20 samples. We show
the cumulative probabilities as a function of the singular values and
find a good agreement; despite the agreement, small singular values are
underestimated in probability for even system sizes. The odd-even effects
are significant and therefore we dedicate the subplots a) and b)
of Fig.~\ref{fig:marchenko} to the even and odd case, respectively.

Based on these results, we extract two more requirements. On the one hand,
we need to sample values according to the distribution $\mathcal{D}_{\mathrm{MP}}$. We draw random
numbers from $\mathcal{U}(0, 1)$ and return the corresponding singular value
for the cumulative probability value. On the other hand, we need to
estimate the truncated singular values; we keep only
$P_{\mathrm{keep}} = \frac{1}{d}$ singular values in the case of an already
saturated bond dimension. We obtain the discretized probabilities according
to the Marchenko-Pastur distribution defined as
\begin{align}
  p^{\mathrm{MP}}(\lambda_{j}) \, , \quad \sum_{j=1}^{N} p^{\mathrm{MP}}(\lambda_{j}) = 1 \, .
\end{align}
The values of $\lambda_{j}$ are sorted ascending and result in
Fig.~\ref{fig:marchenko}.
Finally, the truncation error is given by one sample of singular values
for the given dimension.

\section{Singular value distribution for QFTs                                  \label{sec:sv_distr}}

\begin{figure}[t]
  \begin{center}
    \vspace{0.2cm}
    \begin{overpic}[width=0.75 \columnwidth,unit=1mm]{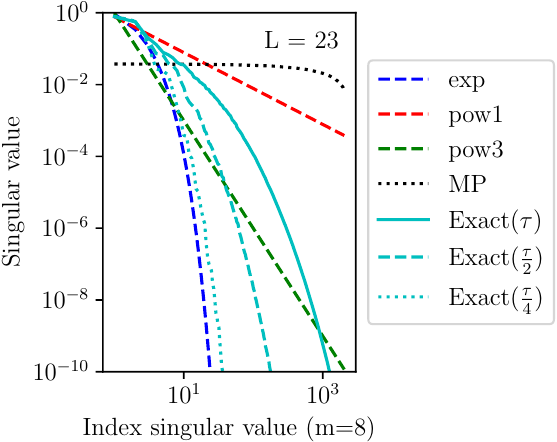}
    \end{overpic}
    \caption{\emph{Singular value distribution for QFTs.} The series of eight
      QFTs are applied to a W-state. The amount of generated entanglement is
      much less than in the Marchenko-Pastur distribution. The cubic power-law
      decay comes closest to the distribution out of our set of distributions.
                                                                                \label{fig:qft_sv_distr}}
  \end{center}
\end{figure}

We evaluate the distribution of the singular values for the quantum Fourier
transformation as we did before for the randomizing circuit.
Figure~\ref{fig:qft_sv_distr} relies on a system size of $L = 23$ as the
previous results for the distribution. Although we start in a W-state
$\ket{W} = (\ket{100\ldots} + \ket{0100\ldots} + \ldots + \ket{\ldots{}001}) / \sqrt{L}$,
a single QFT does not generate much entanglement. Thus, we do a series of QFTs,
i.e., eight in this case. The distribution of the singular values resembles
most a power law with cubic decay, which makes the QFT use case more favorable for classical
emulators.

\end{document}